\documentclass[preprint]{aastex}
\usepackage{natbib}
\newcommand{\Msun}{$M_\sun$}
\newcommand{\mh}{\rm{[M/H]}}
\newcommand{\ot}{NGC~300 OT2008-1}

\shorttitle{The NGC 300 Transient}
\shortauthors{Gogarten et al.}

\begin{document}

\title{The NGC 300 Transient: An Alternative Method For Measuring Progenitor  
Masses}

\author{Stephanie M. Gogarten\altaffilmark{1},
   Julianne J. Dalcanton\altaffilmark{1,2},
   Jeremiah W. Murphy\altaffilmark{1,3},
   Benjamin F. Williams\altaffilmark{1},
   Karoline Gilbert\altaffilmark{1},
   Andrew Dolphin\altaffilmark{4}
}

\altaffiltext{1}{Department of Astronomy, University of Washington,
   Box 351580, Seattle, WA 98195; stephanie@astro.washington.edu}
\altaffiltext{2}{Wyckoff Faculty Fellow}
\altaffiltext{3}{NSF Astronomy and Astrophysics Postdoctoral Fellow}
\altaffiltext{4}{Raytheon; 1151 E.\ Hermans Rd., Tucson, AZ 85706}

\begin{abstract}
  We present an alternative technique for measuring the precursor
  masses of transient events in stars undergoing late stage stellar
  evolution.  We use the well-established techniques of stellar
  population modeling to age-date the stars surrounding the site of
  the recent transient event in NGC~300 (\ot).  The surrounding stars
  must share a common turnoff mass with the transient, since almost
  all stars form in stellar clusters that remain physically associated
  for periods longer than the lifetime of the most massive stars.  We
  find that the precursor of \ot\ is surrounded by stars that formed
  in a single burst between 8--13 Myr ago, 
  with 70\% confidence.
  The transient was therefore likely to be due to a progenitor whose
  mass falls between the main sequence turnoff mass (12 \Msun) and
  the maximum stellar mass (25 \Msun) found for isochrones bounding
  this age range.
  We characterize the general applicability of this technique in
  identifying precursor masses of historic and future transients and
  supernovae (SNe), noting that it requires neither precursor imaging
  nor sub-arcsecond accuracy in the position of the transient. It is
  also based on the well-understood physics of the main sequence, and
  thus may be a more reliable source of precursor masses than fitting
  evolutionary tracks to precursor magnitudes.  We speculate that if
  the progenitor mass is close to 17 \Msun, there may be a connection
  between optical transients such as \ot\ and the missing type II-P
  SNe, known as the ``red supergiant problem.''

\end{abstract}


\section{Introduction}

Energetic and luminous transients mark the final stages of massive
stars' evolution.  These transients include core-collapse explosions
leading to Type II and Ib/c supernovae (SNe), luminous blue variable
(LBV) outbursts, and other more mysterious transients such as the
optical transients in NGC~300 \citep[hereafter \ot, following][] 
{Berger2009} and NGC 6946 \citep[SN~2008S;][]{prieto08b}.
Current stellar evolution models indicate that the mapping between
progenitor stars and the type of
supernova is determined
primarily by the zero age main-sequence mass and the star's
mass-loss history.
The theoretical expectations
are that the most massive stars ($\gtrsim 25$ \Msun) experience
tremendous mass loss at the
end of their lives, expelling their hydrogen and/or helium envelopes
before exploding as SN Type Ib/c; somewhat less massive stars 
($\sim 10$--25 \Msun) are
assumed to lose less mass as they evolve, and explode as SN Type II
with their hydrogen envelopes intact \citep{woosley02,heger03,Limongi2007}.
This picture is supported by observations
\citep{james06,prieto08c,kelly08,anderson08}.
On the other hand,
theoretical expectations for the progenitors of
optical transients such as \ot\ and SN~2008S are lacking.
Empirically, LBVs are expected to be $\gtrsim 20$ \Msun\
\citep{Smith2004,Smith2007} and
estimates for \ot\ and SN~2008S place the progenitor mass somewhere
between 6 and 20 \Msun\ 
\citep{prieto08b,thompson08,Smith2009,Bond2009,Botticella2009}.  
However, it is unclear whether the latter
are SNe resulting from the collapse of O/Ne cores \citep[hereafter referred
to as ``electron-capture SNe'';][]{prieto08b,thompson08,Botticella2009}, an  
extension of the LBV outburst phenomena to lower masses
\citep{Smith2009}, or a new class of transient due to an unknown mechanism.

At this juncture, progress in understanding the origins of these
transients would
be aided by the addition of observational constraints.  The increasing
prevalence of time-domain observations is adding to the number of
transients identified, but the most basic property of the
transients---their stellar mass---remains largely unknown.

One approach to constraining the masses of the stars showing
transients is to use precursor imaging \citep[e.g.,][]{Smartt2009}.  If images of a star exist
before outburst, stellar evolutionary tracks that pass through the color
and magnitude of the likely precursor can then be used to estimate the
star's initial main-sequence mass.
Core-collapse explosions which disrupt the star require
precursor imaging to identify the progenitor,
whereas imaging after the transient is possible for LBVs and other
transients that do not completely disrupt the star.
This second class of objects can be misidentified as SNe and has been
labeled ``SN impostors'' \citep{VanDyk2007}.

Several groups have noted two unusual transients discovered in 2008
with spectra similar to Type IIn SNe, but whose luminosities are much
lower than typical SNe
\citep{stanishev08,bond08,monard08,Bond2009,Berger2009}. These two
transients---SN~2008S and \ot---have no optical precursor, but do show
a very red source in the mid-IR bands covered by IRAC and MIPS on the
{\emph{Spitzer Space Telescope}}
\citep{berger08a,berger08b,prieto08a,prieto08b,Bond2009,Berger2009}.
These two precursors have been interpreted as defining a new class of
heavily dust enshrouded super asymptotic giant branch (AGB) stars,
which may have exploded as electron-capture SNe
\citep{prieto08b,thompson08,Botticella2009}.  Alternatively,
\citet{Smith2009} suggest that SN~2008S is not a core-collapse
supernova, electron-capture or otherwise.  They instead propose that
SN~2008S is a ``SN impostor'' with a super-Eddington wind emanating
from a star with mass 
$\gtrsim 15$ \Msun.

Recent light curve and spectroscopic observations of \ot\
confirm its overall similarity to SN~2008S \citep{Bond2009,Berger2009}.   
The spectra show
evidence for both a wind and fallback of material from a previous
mass-loss event.
However, due to the uncertainties in stellar evolution models,
translating bolometric luminosities to mass estimates results in a
fairly broad range of possible masses for these stars.

While direct imaging of precursors clearly has promise for estimating  
masses, it suffers
from a number of substantial limitations.  First among these is the
requirement that the precursor imaging actually exists. The majority of
past transients have neither pre-existing \emph{Hubble Space Telescope (HST)} imaging, nor
sufficiently accurate astrometry.  The second major limitation is that
even when precursor imaging is available, interpretation of that
imaging depends on the most uncertain stages of stellar evolution
\citep[e.g.,][]{Gallart2005}.  Existing studies typically estimate  
the mass of a
precursor by fitting evolutionary tracks to its color and magnitude.  
However, uncertainties in stellar evolution models related to
rotation, mass
loss, pulsation, internal mixing, the formation of dust in stellar
winds, and convective instabilities in shell-burning layers all
contribute some difficulty in determining the mass of a single 
highly-evolved star (see discussion in \S\ref{sec:models}).

In this paper, we demonstrate an alternative approach using stellar
populations.  Our method is based on the fact that most transient
events occur in the last stages of stellar evolution and within groups
of stars that share a common age and
metallicity.  Thus, even after an individual star has evolved, the
remaining stars still provide information about the age of the stellar
population that hosted the transient, even if the source of the  
transient is no
longer visible.  Once that age is known, one can infer the mass of the
star that exhibited the transient, since it likely corresponds to a star
that has recently turned off the main sequence.  

This method has been used to age-date SNe for which the
surrounding stars could be resolved.  \citet{Efremov1991} identified a
young cluster surrounding SN~1987A and estimated its age as $\sim10$
Myr; this estimate was later refined to $12 \pm 4$ Myr
\citep{Walborn1993} and then $12 \pm 2$ Myr \citep{Panagia2000}.
\citet{vandyk99} studied stellar populations in the vicinity of a
number of SNe and were able to estimate progenitor masses in a few
cases.
\citet{Maiz-Apellaniz2004} and \citet{Wang2005} estimated the age and
mass of SN~2004dj
using not resolved stars in its originating cluster but by finding the
combination of spectral energy distributions (SEDs) which best fit the
integrated cluster light.  They arrived at age estimates of 13.6 Myr and
20 Myr, respectively, which correspond to progenitor masses of 15 and
12 \Msun.  \citet{Vinko2009} fit isochrones to the color-magnitude
diagram (CMD) of
resolved stars in this cluster and placed the age at $\sim10$--16 Myr,
corresponding to a progenitor mass of $\sim 12$--20 \Msun.
\citet{Crockett2008} attempted to date SN~2007gr using SED fitting, but
found two possible solutions at $7 \pm 0.5$ Myr and 20--30 Myr.

Our method takes advantage of the fact that most stars form in stellar
clusters with a common age ($\Delta
t \lesssim 3$--5 Myr) and metallicity.  Indeed, over 90\% of stars
form in rich clusters containing more than 100 members with
$M > 50$ \Msun\ \citep{Lada2003}.  The stars which formed in a common
event remain spatially correlated on physical scales up to
$\sim 100$ pc during the 100 Myr lifetimes of 4 \Msun\ stars, even
if the cluster is not gravitationally bound \citep{Bastian2006}; we have
confirmed this expectation empirically in \citet{Gogarten2009}.  Thus,
it is reasonable to assume that most young stars within approximately a hundred
parsecs of a massive-star transient are coeval.  The age of a transient's host
stellar population can then be recovered from the
CMD of the surrounding stars, using well-established methods
for deriving the star formation history (SFH).

Herein, we analyze the specific case of \ot, by
extracting stars from a small aperture around the location of the
transient, using photometry from the ACS Nearby Galaxy Survey Treasury
\citep[ANGST;][]{Dalcanton2009}.  Using the methods described in
\citet{Williams2009} and \citet{Gogarten2009}, we solve for the SFH in
the region of the transient.  We find a
very well-defined burst of star formation that occurred 8--13 Myr ago.
The burst allows us to accurately age-date the evolving stellar
population which produced the transient.  The mass of the transient
can then be linked to the masses of stars that have recently turned
off the main sequence
but have not yet exploded,
which in this case is 
12--25 \Msun.
We will show that with this method, it is possible to establish firm limits
on the
mass of the precursor,
using only imaging taken {\emph{after}} the transient itself.

The outline of the paper is as follows.  In \S\ref{sec:data} we
discuss the stellar photometry used to analyze the age of the
transient.  In \S\ref{sec:sfh} we discuss the recovery of the SFH.  In
\S\ref{sec:mass} we calculate the resulting stellar mass of the main
sequence star which evolved to produce the transient.  In
\S\ref{sec:discussion} we discuss the implications of our results,
strategies for expanding the use of
this method, and possible limitations.  We conclude with
\S\ref{sec:conclusions}.

\section{Data and Photometry}
\label{sec:data}

The location of \ot\ \citep[$\alpha = 00^{\rm h}54^{\rm m}34.552^{\rm s}$, $\delta =
-37\arcdeg38\arcmin31.79\arcsec$ (J2000),][]{berger08b} is
contained both in one of our ANGST
fields observed 2006 November 8, and in an archival observation taken
2002 December 25.  The ANGST data had exposure times of 1488s in
$F475W$, 1515s in $F606W$, and 1542s in $F814W$.  The archival data had
exposure times of 1080s in $F435W$ and $F555W$, and 1440s in $F814W$.
Each of these observations was
split into two exposures, which were calibrated and flat-fielded using
the standard \emph{HST} pipeline.  Photometry was done using DOLPHOT,
a modified version of HSTphot \citep{Dolphin2000} optimized for ACS.
Details of the photometry and quality cuts used for the ANGST sample
and archival data are given in \citet{Williams2009} and
\citet{Dalcanton2009}.  We use only the highest-quality photometry for
this analysis.  Although the ANGST data was in fact taken before the
transient occurred, there is no star visible at the location of the
transient \citep{Bond2009}, and thus the precursor was not visible at
optical wavelengths.

Artificial stars, used to characterize the completeness of
the photometry, were inserted and
detected using DOLPHOT.  We have approximately 200,000 artificial
stars for each field in the region in which both the archival and
ANGST fields overlap.  We use fake stars only for the overlap region
so the effects of crowding will be identical for the two fields.

\begin{figure}
\plottwo{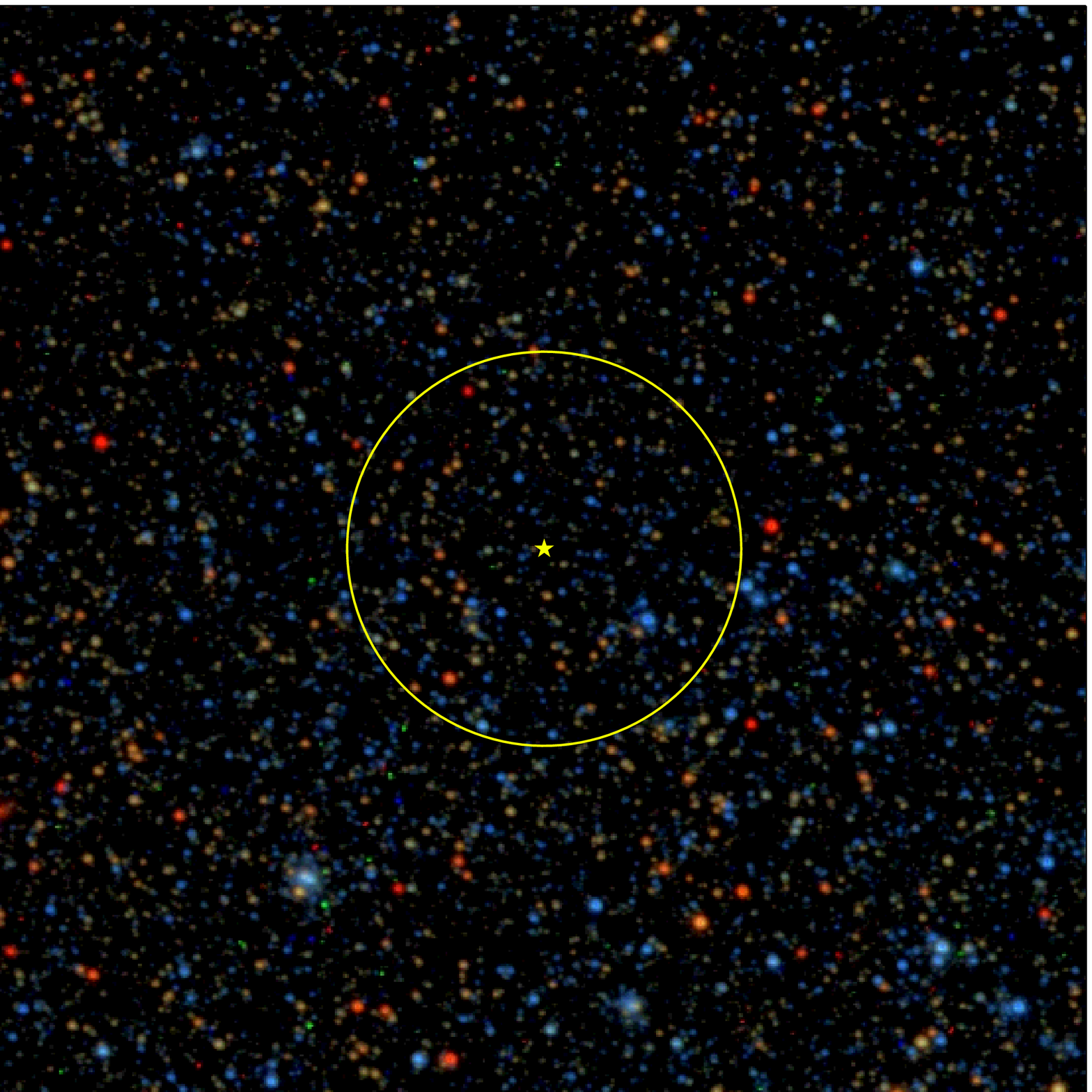}{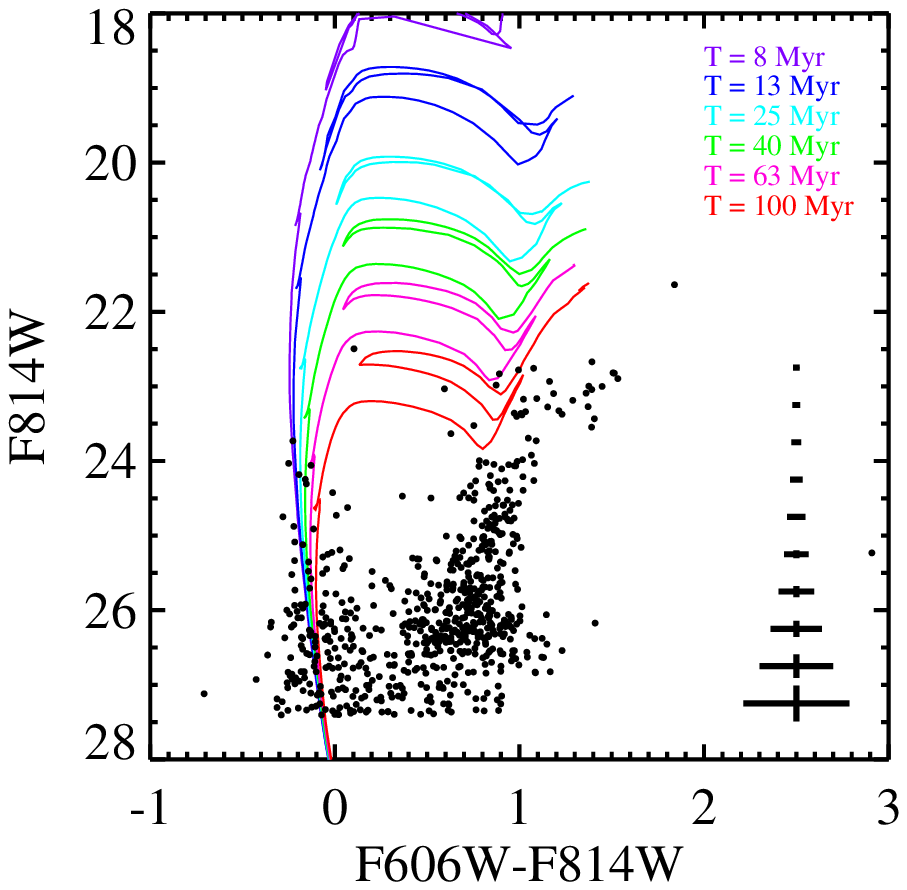}
\caption{\label{fig:cmd}
   Left: subsection of the 3-color ($F475W, F606W, F814W$) 2006 ACS  
image of
   NGC~300.  The location of the transient is marked with a star, while
   the circle indicates the $5\arcsec$ radius from which we selected
   stars.  North is up and east is to the left.
   Right: CMD of stars in the selected region.
   Photometric errors as a function of $F814W$ magnitude are shown on
   the right.
   Isochrones corresponding to the edges of our age bins for SFH
   derivation are plotted.  All isochrones are from \citet{Marigo2008}
   and  have $\mh = -0.4$, $A_V = 0.1$, and $m-M = 26.5$.}
\end{figure}

We select stars in a $5\arcsec$ radius around the transient location
(Figure~\ref{fig:cmd}).  This translates to $\sim 50$ pc at the
distance of NGC~300 (2.0 Mpc), which is a compromise between including
as many coeval stars as possible while limiting contamination 
(see \S\ref{sec:region} for a discussion of the choice of
selection radius).
The CMD shows a truncated main
sequence, indicating that star formation in this
region terminated more than $\sim 5$ Myr ago.  However, the CMD shows
no core helium burning stars, suggesting that there was no significant
star formation during the interval between 25 and 300 Myr ago.  Thus,
even without a detailed analysis, the CMD alone indicates that the
transient was due to a massive star that formed during a
burst of star formation between $\sim 5$ and 25 Myr ago.
In the next section, we analyze
the SFH in depth, giving tighter constraints on the age.

\section{Analysis}

\subsection{Star Formation History}
\label{sec:sfh}

Overlaying isochrones on a CMD is the traditional method of
determining the age of a young stellar population; however, this method
relies on only a few main-sequence stars to establish age.
In contrast, fitting
the entire CMD uses the full wealth of information and can give
tighter age constraints.
Comparing the observed CMD to a set of model CMDs to
derive the star formation history (SFH) is a well-established technique
\citep{Hernandez1999,Gallart1999,Holtzman1999,Dolphin2002,Skillman2003,Harris2004,Gallart2005}.
While there are many different codes available, the basic procedure is
the same for all: stellar evolution models are used to predict the  
properties
of stars of different masses for a range of ages and metallicities.
 From the predicted luminosity and temperature, the magnitudes of the
stars are determined for a given filter set.  For each age and
metallicity, stars are placed on a synthetic CMD in proportion to an
assumed initial mass function (IMF).  These CMDs are then linearly
combined, with distance and extinction either fixed or included as
additional free parameters, until the best fit to the observed CMD is
found.  The ages and metallicities of the CMDs that went into the best
fit tell us the ages and metallicities of the underlying stellar
population, while the weights given to the CMDs provide the SFR at
each age.

To derive the SFH for \ot, we use MATCH, which
finds the maximum-likelihood fit to the CMD \citep{Dolphin2002}.
We assume an IMF with a slope of
-2.35 \citep{Salpeter1955} and a binary fraction of 0.35.  MATCH only
allows a single value for the slope of the IMF, but given that 
our CMD only includes stars with masses $>1$ \Msun, adopting a single
Salpeter slope is likely to be a valid assumption.
Synthetic CMDs are constructed from the theoretical isochrones of
\citet{Marigo2008} for ages in the range 4 Myr--14 Gyr.
The isochrones younger than $\sim6\times 10^7$~yr
  were adopted from \citet{Bertelli1994}, with transformations to
  the ACS system from \citet{Girardi2008}.
Age bins are spaced logarithmically since the CMD
changes much more rapidly at young ages than at old ages.
The edges of the age bins shown in this paper are 4 Myr (the age of
the youngest isochrone), 8, 13, 25, 40, 63, and 100 Myr.
Additional age bins go back to 14 Gyr.

Metallicity is constrained to increase with time within the range
$-2.3 < \mh < 0.1$.  As additional free parameters, the distance
modulus is allowed to vary in the range $26.3 < m-M < 26.7$ and
extinction can vary in the range $0.05 < A_V < 0.50$.  Also,
up to 0.5 mag of differential extinction may be applied to young stars
($< 100$ Myr).  The \citet{Schlegel1998} value for Galactic extinction
is $A_V = 0.042$ in the line of sight to NGC~300, but we expect  
the
total value to be higher due to local extinction within NGC~300 itself.

Comparisons to model CMDs were carried out within bins of width 0.1  
mag in color and 0.2 mag in
magnitude.  For a small number of stars, choosing bins that are too
small results in so few stars in each bin that the accuracy of the
fitting suffers. Our choice of bin size reduces this problem while
ensuring that the number of bins in the CMD is substantially larger
than the number of free parameters in the fit.

Completeness and photometric biases were accounted for by including  
the results of the
artificial star tests. We supplied MATCH with the input and output
magnitudes of the artificial stars and whether they were detected
above the quality cuts of our photometry.  We include only the portion
of the $F606W+F814W$ CMDs that were more than 50\% complete, which  
corresponds to magnitudes
of $F606W < 27.9, F814W < 27.0$ for the ANGST data and $F555W < 27.6,
F814W < 27.0$ for the archival data.  We also recovered the SFH
using the $F435W/F475W+F814W$ filter combinations and found consistent
results, but since the
depth in the $B$ equivalent filters is less than in the $V$ equivalent
filters ($F555W/F606W$), the
error bars on the resulting SFHs were considerably larger.

\begin{figure}
\plottwo{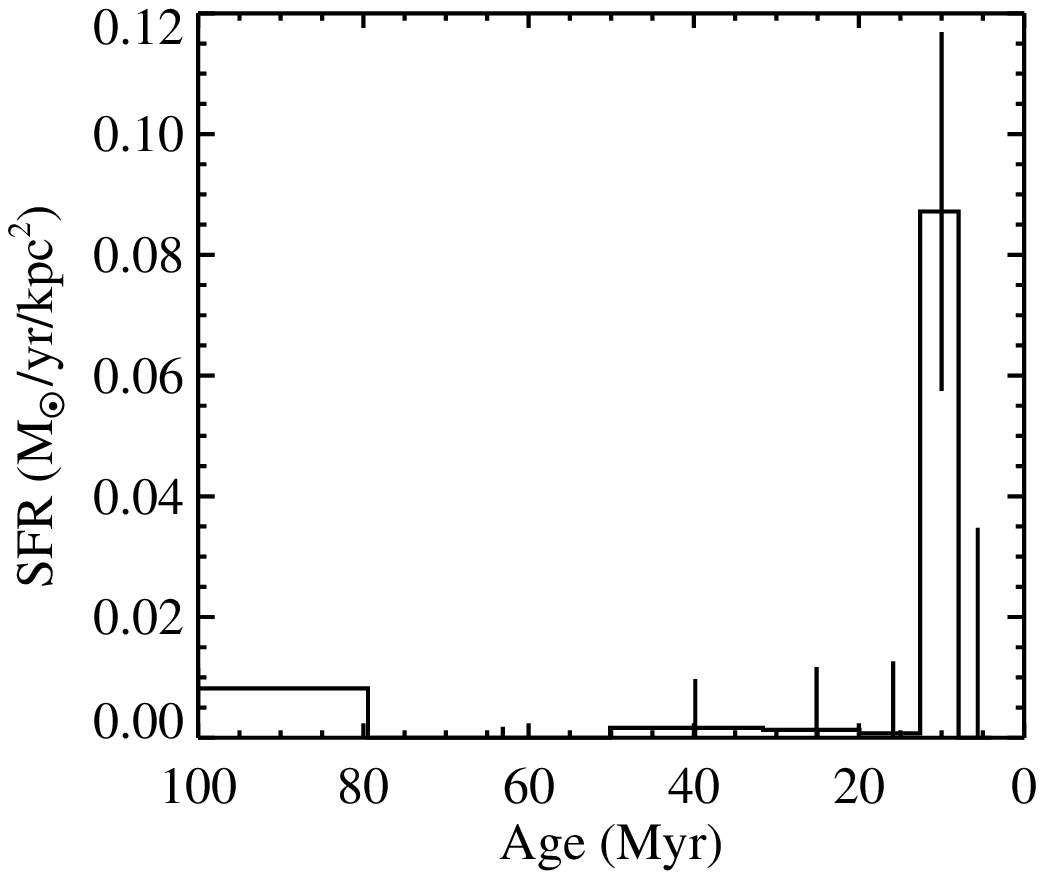}{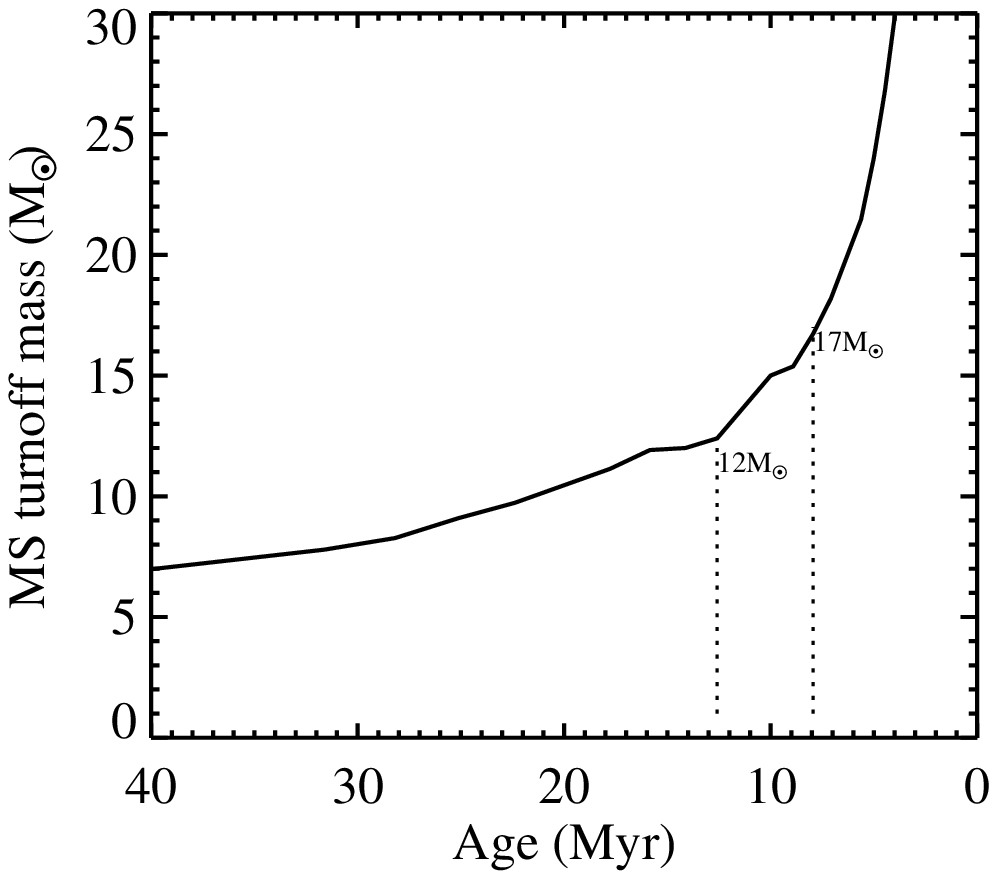}
\caption{
   \label{fig:sfh}
   Left: recent star formation history ($<100$ Myr) of the selected
   region in NGC~300, showing a significant star formation event at
   8--13 Myr.  SFH was found by comparing the observed ANGST CMD with
   synthetic CMDs based on theoretical isochrones.  Error bars are the
   quadrature sum of the uncertainties from distance and extinction
   and the 68\% confidence interval from Monte Carlo simulations.
   Right: main-sequence turnoff mass vs.\ age \citep{Marigo2008}.  The
   boundaries of the star formation event and corresponding stellar
   masses are marked.  
These masses represent the lower limits on the mass of the transient.
}
\end{figure}

The resulting SFH of this region over the past 100
Myr is shown in Figure~\ref{fig:sfh}.  The SFH shows a single star  
formation
event in the range 8--13 Myr ago,with no significant additional star
formation in the age range of 0--100 Myr.
This isolated burst is the likely source of the
progenitor.  The time of the star formation event agreed between the
ANGST-derived and archival-derived SFHs, with the amplitudes of the
star formation events agreeing within the error bars.
We note that many of the fainter stars in this region were formed at a
low SFR over the past 14 Gyr; however, we plot only the past 100 Myr
to highlight the event that likely produced the progenitor.

Error bars in Figure~\ref{fig:sfh} are the quadrature sum of the
errors from uncertainties in distance and extinction and
the 68\% confidence interval from Monte Carlo tests, which assess
uncertainties due to limited numbers of stars.  Monte Carlo
simulations were run by sampling stars
from the best fitting model CMD determined by MATCH.  These stars were
then given as the input to MATCH, and the resulting SFH was compared
to the SFH from which they were drawn.  This process was repeated 100
times, and the scatter in difference between the input and output SFHs
was incorporated into the error bars in our reported SFHs for each
region.  These trials indicated that the residual recent star  
formation at
ages older than the burst was unlikely to be real, and that the SFH
within the past 50 Myr could be explained entirely with a single burst.
Monte Carlo simulations assess uncertainties
due to Poisson sampling of underpopulated regions in the CMD, but they are
not sensitive to systematic uncertainties in the models themselves.
Hence, our error bars for the SFH do not include model uncertainties,
but should be viewed as the range of possible values for this
particular set of models.  We discuss model uncertainties further in
section \S\ref{sec:models}.

Extinction was found to be $A_{F606W} = 0.09 \pm
0.04$.
\citet{Gieren2005} find a total reddening for NGC~300 of $E(B-V) =
0.096 \pm 0.006$ mag, which corresponds to an extinction of $A_V =
0.3$ assuming the extinction law of \citet{Cardelli1989}.
Given that the extinction in NGC~300 varies with position
\citep{Roussel2005}, it is not surprising that the value for this
particular region is smaller.
\citet{Berger2009} estimate the extinction of the transient itself from
SED fitting (after much of the dust surrounding the
progenitor may have been destroyed) and find $A_V = 0$ with a 1$\sigma$
upper limit of $A_V < 0.6$.  Our value is consistent with this
estimate also, although we note that the value we derive is the
average extinction of the stars around the transient and not of the
transient itself, which would include the circumstellar dust that
prevented the detection of the progenitor at optical wavelengths.
\citet{Bond2009} find from their spectroscopy that the transient
itself could be reddened by as much as $E(B-V) \simeq 0.4$,
corresponding to an extinction of $A_V \simeq 1.2$.

The best-fit distance modulus was found to be $26.41 \pm 0.09$, which agrees with
the value of 26.50 found using the tip of the red giant branch for the
entire ACS field \citep{Dalcanton2009}, as well as
previous distance estimates \citep{Butler2004,Gieren2005,Rizzi2006};
however, this is by construction given the limits
placed on the distance modulus.

The current metallicity returned by MATCH is $\mh = -0.44 \pm 0.17$,
which is in agreement with the value of $\mh = -0.27 \pm 0.15$ at this
galactocentric radius (2.7 kpc) from the observed metallicity  
gradient of
\citet{Kudritzki2008}.  
To test the effect of metallicity on our results, we fixed the
metallicity at $\mh = -0.7 \pm 0.1$ when deriving the SFH.  The SFH  
still
shows a burst at 8--13 Myr ago, though with an increased probability of
the burst beginning earlier (20 Myr ago), presumably due to redder
stars being interpreted as older rather than more metal-rich.
Forcing the stars to be at solar metallicity results in a burst that
is too young, 4--10 Myr ago, but that scenario is clearly inconsistent
with the measured metallicity at this radius.

\subsection{Mass of Progenitor}
\label{sec:mass}

The relationship between stellar age and the main-sequence turnoff
mass is plotted in Figure~\ref{fig:sfh} \citep{Marigo2008}.  From
this plot we can infer the progenitor mass for a stellar population of
a given age.  The star formation event which gave rise to the transient
happened between 8 and 13 Myr ago, which corresponds to a 
main sequence turnoff mass that falls in the
mass range
of 12--17 \Msun.
Assuming that the transient was due to an evolving star, the transient
precursor had a mass higher than this lower limit.

To determine the uncertainty in our 
turnoff
mass estimates, we ran a series of
Monte Carlo tests using artificial star formation bursts of varying  
amplitudes and
durations and measured how well these SFHs
could be recovered.  We varied the amplitude
of the bursts (i.e. SFR or number of stars formed in a constant time
period), the start and end times of the burst (for varying duration),
and the age of the burst for similar durations.  Each simulation was
repeated 50 times.

\begin{figure}
\plotone{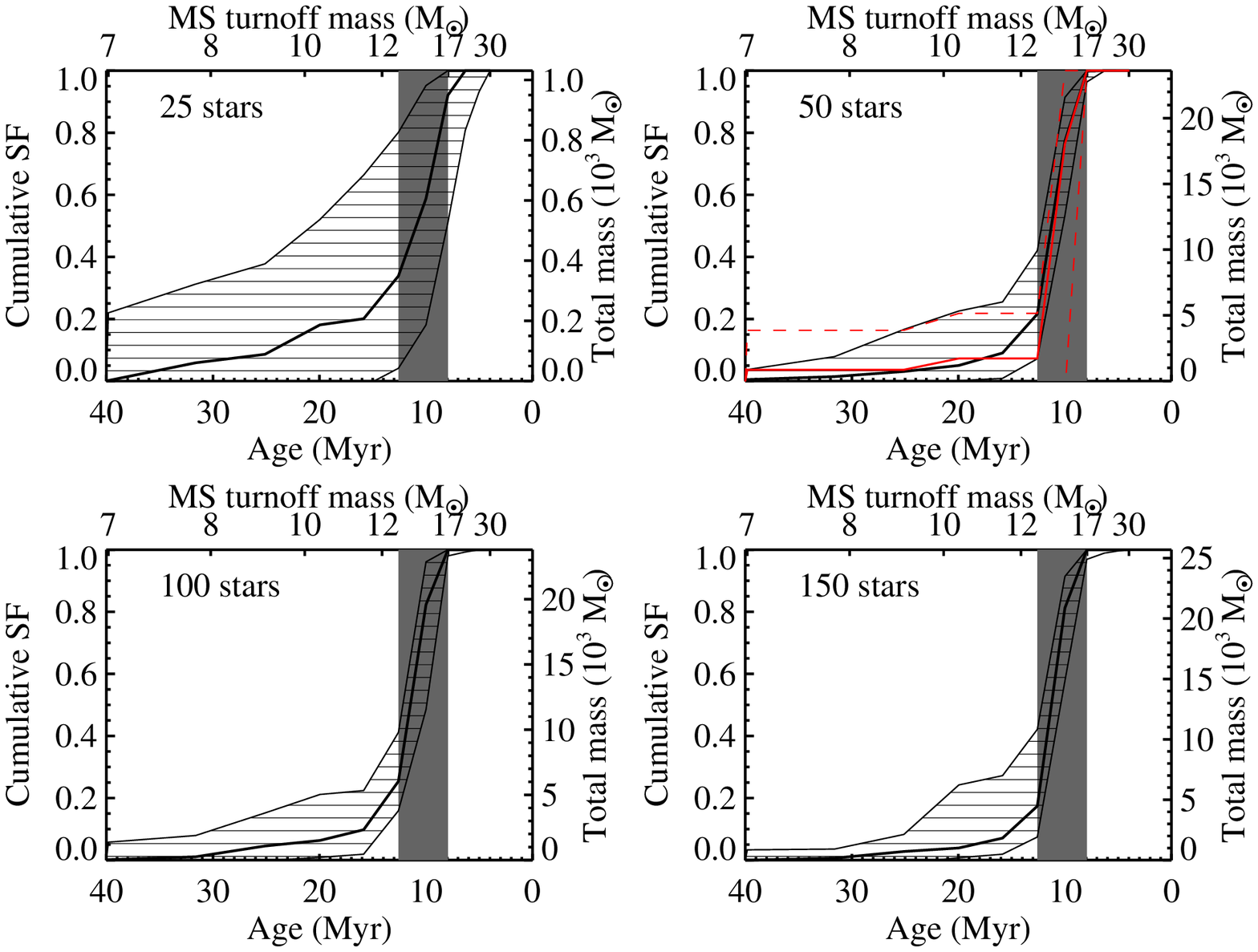}
\caption{\label{fig:burst_nstars}
   Simulated bursts of age 8--13 Myr and varying amplitudes (gray shaded
   regions) and derived SFHs.
   The cumulative distribution functions show the total star formation
   summed from older ages to younger ages.
   The central black line is the median value of the Monte Carlo  
tests, and
   the surrounding hashed region is the 1$\sigma$ error range.
   Derived SFH and errors for \ot\ are in red.
   The number of stars refers to upper main-sequence stars ($F606W -
   F814W < 0.3$, $F606W < 26$).}
\end{figure}

The amplitude trials were set so as to produce a certain number of  
upper main
sequence stars (defined as $F606W - F814W < 0.3$ and $F606W < 26$).
Trials were run with numbers of approximately 25, 50, 100, and 150
(exact numbers varied slightly with
each trial, since CMDs were created by sampling from the fixed SFH).
The burst duration was set to 8--13 Myr to match the
derived SFH of \ot.
Results are shown in Figure~\ref{fig:burst_nstars}.
If each burst were perfectly recovered by MATCH, the cumulative
distribution would go from 0 to 1 within the gray shaded region which
shows the burst duration.
All of the trials have a tail to older ages
and smaller masses, indicating the tendency of MATCH to interpret
stars as older than they really are, but this accounts for less than
40\% of the total star formation in all cases with 50 or more upper
main-sequence stars.
  The star burst with $\sim 50$ upper main-sequence stars is the  
closest match
to the derived SFH, which had 55 upper main-sequence stars.
We see that the burst is not well constrained with only 25 upper main
sequence stars, but it is at 50 and above.  We conclude that at least
50 upper main-sequence stars are required to use this method.

\begin{figure}
\plotone{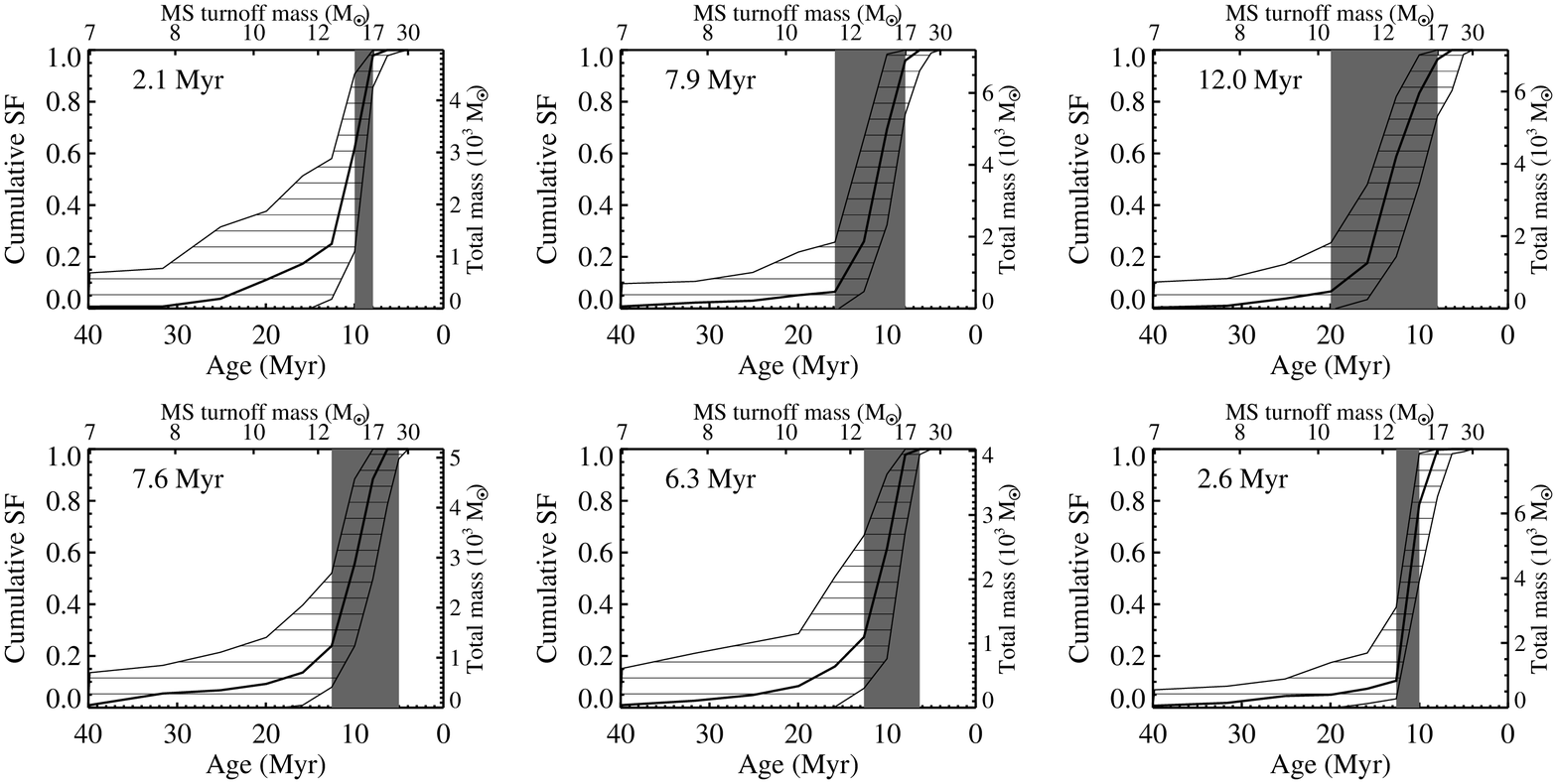}
\caption{\label{fig:burst_duration}
   Simulated bursts with $\sim 50$ upper main-sequence stars and
   varying durations (gray shaded regions) and derived SFHs.
   The central black line is the median value of the Monte Carlo  
tests, and
   the surrounding hashed region is the 1$\sigma$ error range.
   Upper row: constant end time (bursts start at 10, 16, and 20
   Myr ago and end at 8 Myr); lower row: constant start time (bursts  
start
   at 13 Myr ago and end at 5, 6, and 10 Myr).
   Burst duration is labeled on each plot.}
\end{figure}

We next set all duration trials to produce approximately 50 upper  
main-sequence
stars, and then varied the starting and ending times of the burst.
The end time trials began at 13 Myr ago and ended at 5, 6, and 10  
Myr, and the
start time trials began at 10, 16, and 20 Myr ago and ended at 8 Myr
  (Figure~\ref{fig:burst_duration}).
These can be compared with the
8--13 Myr burst from Figure~\ref{fig:burst_nstars}.
We see that very short duration bursts (2--3 Myr) are not
able to be recovered, with the resulting SFHs extending to
significantly younger and older ages.  Longer bursts are 
better
constrained, especially at the upper 
turnoff mass limit.
Despite the long
tail to older ages, this actually represents a small range in
absolute 
turnoff
masses, as the 7--10 \Msun\ range stretches over almost 20 Myr.

\begin{figure}
\plotone{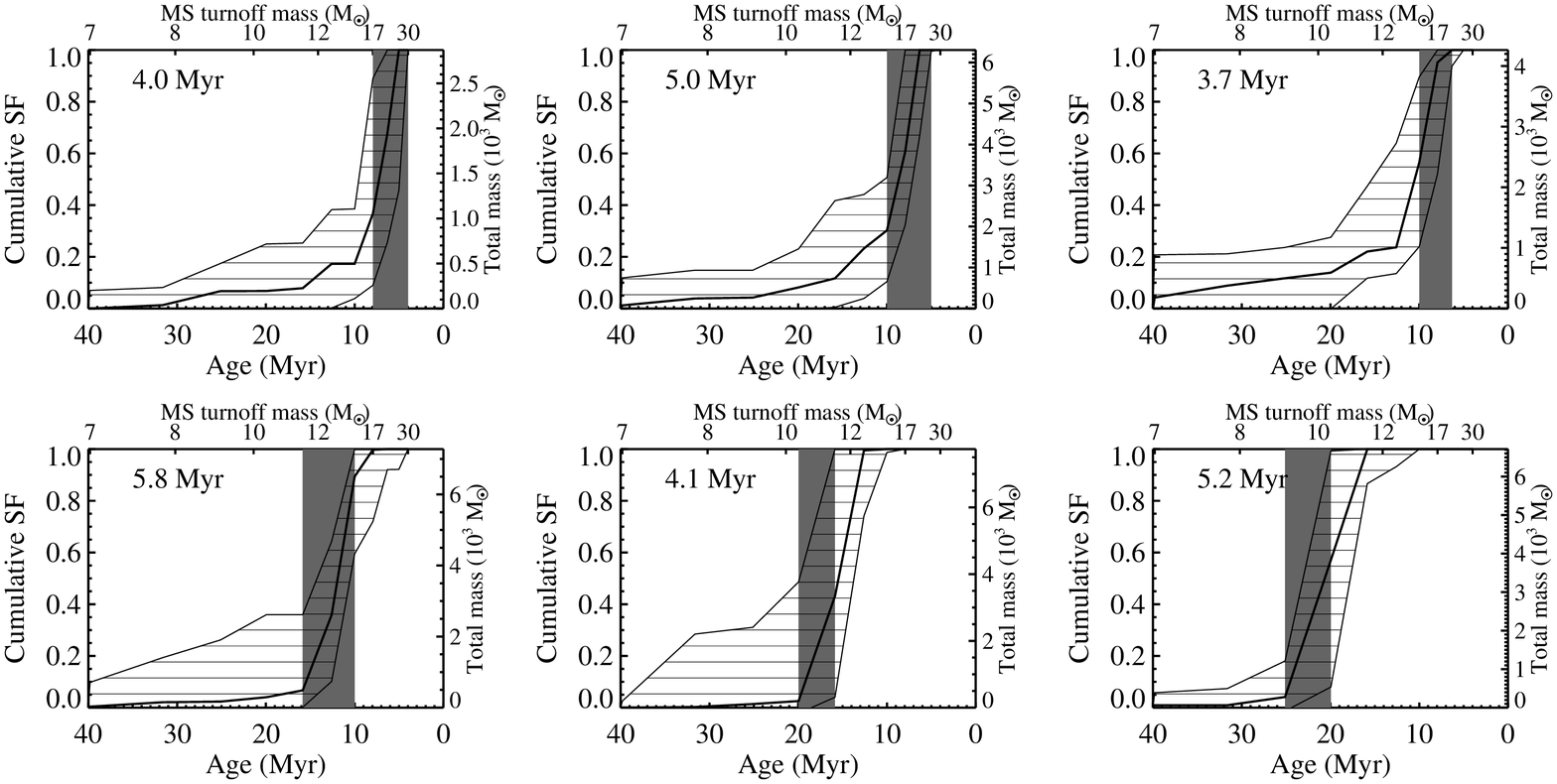}
\caption{\label{fig:burst_age}
   Simulated bursts with $\sim 50$ upper main-sequence stars and
   varying ages (gray shaded regions) and derived
SFHs.
   The central black line is the median value of the Monte Carlo  
tests, and
   the surrounding hashed region is the 1$\sigma$ error range.
   Upper row: burst ages are 4--8, 5--10, 6--10 Myr.  Lower row: burst
   ages are 10--16, 16--20, and 20--25 Myr.
   Burst duration is labeled on each plot.}
\end{figure}

Since we used logarithmic time binning, we could not fix a constant
linear time duration for a variety of ages, so instead we chose the  
time bins
which gave durations as close as possible to the derived duration of
4.6 Myr.  Figure~\ref{fig:burst_age} shows the results of these
trials, with start times of 4, 5, 6, 10, 16, and 20 Myr ago.
 From these plots we conclude that bursts of this length
are more easily recovered at more recent times.
Bursts that ended more than 10 Myr ago are frequently interpreted as
ending more recently.  We speculate that this problem may be related
to poor sampling.  If upper main-sequence stars are present, the CMD
must be young, but the absence of these stars could be due to chance at
the low levels of star formation that we are probing.
Bursts with very recent end times show no tails to more recent ages
because the youngest isochrones in our models are at 4 Myr.

We use a subset of these tests to assess the uncertainty in our mass
estimate.  The question we are asking is whether a burst that started
earlier or ended later might be interpreted by MATCH as the 8--13 Myr
burst we derived, and with what probability this might occur.
For the start time variation, we define a ``matching'' burst as one
that has a SFR $< 10^{-5}$ \Msun\ yr$^{-1}$ for ages older than
13 Myr and a SFR $> 10^{-4}$ \Msun\ yr$^{-1}$ at some point in the
8--13 Myr range.  These cutoff values are taken from the derived
SFH of \ot, which has SFR $> 10^{-4}$ \Msun\ yr$^{-1}$ during the  
burst and
$< 10^{-5}$ \Msun\ yr$^{-1}$ at other times.  For the end time
variation, we likewise define a ``matching'' burst as one that has a
SFR $< 10^{-5}$ \Msun\ yr$^{-1}$ for ages younger than
8 Myr and a SFR $> 10^{-4}$ \Msun\ yr$^{-1}$ at some point in the
8--13 Myr range.  We ran 50 additional Monte Carlo tests for these
bursts, for a total of 100, but the results remained the same as for
the original 50 runs testing the recovery of burst duration.

Of the
bursts that started earlier (8--16 Myr and 8--20 Myr ago, top center and
right panels in Figure~\ref{fig:burst_duration}), we found that the
probabilities of recovering a single 8--13 Myr burst were 40\% and 22\%,
respectively.  This means that if we derive an 8--13 Myr burst for an
observed CMD, we can be 60\% sure that the burst started after 13 Myr
($M_{prog} > 12.4$ \Msun) and 78\% sure that it started after 16 Myr  
($M_{prog} > 11.9$ \Msun).
Of the bursts that ended later (6--13 Myr and 5--13 Myr ago, bottom left
and center panels in Figure~\ref{fig:burst_duration}), we found
probabilities of 30\% and 26\%, respectively.  Thus, we can be 70\%
certain that the burst ended by 8 Myr ago ($M_{prog} < 17$ \Msun) and  
74\% certain
that it ended by 6 Myr ago ($M_{prog} < 20$ \Msun).  Based on these  
tests, we
report our uncertainty on the 12--17 \Msun\ 
turnoff
mass estimate as 70\% ($\sim 1 \sigma$ in a Gaussian distribution),
setting the lower limit to the transient precursor mass.

At the higher mass end, the mass of the precursor is limited by
the most massive post-main sequence star that is expected to exist
for a burst of star formation between 8--13\,Myr ago.  As we show
in Figure~\ref{fig:iso_compare}, isochrones for 8--13\,Myr old
bursts show no stars currently more massive than $\sim$16--25
\Msun, with $\sim$1 \Msun variation between different stellar
models.  If these models are correct, then the progenitor mass
lies above 12--17 \Msun\ but below 16--25 \Msun.

To test the effect of varying the IMF and binary fraction on our
results, we generated artificial SFHs with a burst from 8--13 Myr,
distance modulus of $m-M=26.5$, extinction of $A_{F606W}=0.1$, and
metallicity of $\mh=-0.4$, matching what we derived for the region
surrounding \ot.  We varied the IMF from $\alpha=-2.0$ to
$\alpha=-2.7$ for a binary fraction of 0.35, and varied the binary
fraction from 0.2 to 0.5 with a Salpeter IMF of $\alpha=-2.35$.  We ran MATCH
on each of these models assuming a Salpeter IMF and binary fraction of
0.35, as in our original SFH derivation, to see if the age of the
recovered burst changed with the input IMF or binary fraction.  We
found that the only changes were in the amplitude of the burst, as
expected since changing the IMF changes the normalization of the SFH,
and
that in each case MATCH still recovered a burst of 8--13 Myr.  Hence,
even if the IMF or binary fraction is different from our assumed
value, it should not affect our age or mass estimates.

\section{Discussion}
\label{sec:discussion}

\subsection{Implications for SN~2008S/\ot-type Transients}

Both \ot\ and the similar transient SN~2008S
have very highly obscured progenitors that are bright point sources at
mid-infrared wavelengths but are undetectable in the optical.
The spectrum of SN~2008S is similar to a SN IIn spectrum,
but many scenarios, including the outburst of a star, are consistent
with the low velocities implied by the narrow lines \citep{Dessart2009}.
To date,
two possible explanations have been put forward to explain this class
of luminous transient.
\citet{thompson08} suggest that the progenitors were heavily dust
enshrouded AGB stars of mass $\sim 10$ \Msun\ with a photospheric  
radius in
the mid-IR of
a few hundred AU, and that the transients may have been
electron-capture SNe.  
\citet{Botticella2009} also endorses the electron-capture SN
explanation for these transients, with a mass estimate of 6--8 \Msun\
for SN~2008S.
Alternatively, \citet{Smith2009} argue that the
transient may be due to a super-Eddington wind being driven from a
$\gtrsim20$ \Msun\ star.
The SED and light curve of M85~OT2006-1 are
also very similar to \ot\ and SN~2008S \citep{Kulkarni2007}, with a
mass estimate on the low end, $\lesssim 7$
\Msun\ \citep{Ofek2008}.

The 12--17 \Msun\ main-sequence
turnoff
mass we have derived
as a lower limit
for the precursor is in between the mass ranges suggested for both the
electron-capture SN in an AGB star \citep[6--11
\Msun,][]{thompson08,Botticella2009} and the wind mechanism suggested
by \citet{Smith2009}.  Although \citet{Smith2009} suggests a rough
lower limit of 
$\sim15$
solar masses for the super-Eddington wind,
there are no firm theoretical arguments that would exclude lower
masses.  Given both the uncertainty in these mass ranges and the
possibility that our data is consistent with a mass as high as 
25
\Msun\ or as low as 11 \Msun, we cannot at present rule out either
option.

Our mass estimate is consistent with the $\sim 10$--15 \Msun\ estimate
of \citet{Bond2009} from \emph{Spitzer} mid- to far-infrared fluxes.
They infer an outflow from the spectrum of the transient, and propose
an OH/IR star which erupted and cleared its surrounding dust envelope.
\citet{Berger2009} find evidence of infall as well as outflow in their
high-resolution spectra.  
Using a similar interpretation to \citet{Smith2009}, they suggest a
blue supergiant or Wolf-Rayet
star and give a more generous mass estimate of
$\sim 10$--20 \Msun, also fully consistent with our derived SFH.

\subsection{The Red Supergiant Problem}

The progenitor mass for \ot\ reported in this paper has
interesting implications for stellar evolution theory.
Prior to SN~2008S and \ot, episodic mass
loss evident in LBV outbursts and SN~IIn was associated only with
the most massive ($>20$ \Msun) stars \citep{Smith2004,galyam07,Smith2007}.
Whether \ot\ is an electron-capture SN
\citep{prieto08b,thompson08} or a non-disruptive outburst analogous to
LBV eruptions \citep{Smith2009,Berger2009}, the narrow emission lines are
a signature of significant mass loss at speeds lower than a typical SN.
In either case, the association of a progenitor mass as low as 
$\sim$15
\Msun\ with an outburst and implied episode of mass loss is a new and
somewhat unexpected result.
Perhaps the only obvious analog is  
SN~1987A, which was a $\sim 20$ \Msun\ blue supergiant just before
explosion
\citep{woosley1987,walborn87,woosley88a,woosley88b,arnett91}.
Its progenitor was
likely a red supergiant (RSG) $\sim30,000$ yr earlier \citep{fransson89},
and may have made the transition via a LBV-like eruption
\citep{Smith2007}.  Might \ot\ be the signature of a similar
transition?

Whether significant mass-loss events are common
for stars $<20$ \Msun\ has important
consequences for our understanding of stellar evolution, especially
during the last stages.
If the cumulative effect of events like \ot\ is to
expel a significant fraction of the hydrogen envelope, then we know
that not all $\sim8$--20 \Msun\ stars experience such events, since
the most common type of SN, SN~II-P, originate from stars in this mass
range.
However, \citet{Smartt2009} have
identified a potential lack of SN~II-P progenitors more massive than
$16.5 \pm 1.5$ \Msun. 
RSGs, the progenitors of SNe~II-P, should be common up to a maximum
mass of at least $\sim25$ \Msun,
since more massive stars will become Wolf-Rayet stars
  \citep[e.g.,][]{Crowther2007}.  
Although statistically some SN progenitors of masses 17--25 \Msun\
should have been in the volume-limited sample of \citet{Smartt2009},
they report with $2.4\sigma$ confidence that RSGs in this mass range
are not exploding as SNe~II-P as expected.
\citet{Smartt2009}
consider several resolutions to this problem, including the idea that
stars in this mass range explode as another type of SN, but they
favor the idea that the minimum progenitor mass for black hole formation is
lower than expected, resulting in failed or weak SNe for this mass range.

Intriguingly, the progenitor mass we find for \ot\ is consistent with the
mass range that defines the onset of the SN~II-P deficit, and the
event itself could signify the transition away from a RSG phase.  
We consider whether stars in the
mass range $\sim17$--25 \Msun\ experience optical transients (OTs) and
episodic mass-loss,
expelling much or some of their hydrogen
envelopes and aborting a connection to SN~II-P.
Using the observed rate of SNe~II-P and assuming the Salpeter IMF, we
can estimate the minimum number of \ot-like transients necessary
to explain the RSG problem.  If we assume that a minimum of one
OT per star is necessary to keep it from exploding as a SN~II-P,
integrate the IMF
from 17 to 25 \Msun, and compare this to the IMF integrated
from 8 to 17 \Msun, this gives us a minimum OT-to-SN~II-P ratio
of 0.23 to explain the lack of high mass SN~II-Ps.  

\citet{Smartt2009} report 54 SN~II-P in their
volume limited ($< 28$ Mpc) survey during a span of 10.5 years.  The
deficit of higher mass SN~II-P progenitors suggests that (in
steady state) at least $\sim 12$ additional stars within the same
volume must have shed some of their hydrogen envelopes in an OT during the
duration of the survey; an even higher rate of OTs may be expected if
multiple mass loss events are needed to fully strip the hydrogen
envelope.  Three similar  transients were reported between 2006
and 2008, one in 2006 and two in 2008 \citep{Berger2009}.  They are
\ot\ at $\sim 2$ Mpc, SN~2008S at $\sim 4$ Mpc, and M85~OT2006-1
at $\sim 17$ Mpc.  If we take this as a rough rate of OTs, then the  
number
of OTs over the $>10$ yr span of the \citet{Smartt2009} survey is $\sim11$,
and within a smaller survey volume.  Therefore, the observed rate of
OTs is not inconsistent with this type of event being sufficient to
explain the paucity of high mass SN~II-P progenitors between
17 and 25 \Msun.  However, given the uncertainties and small sample size
in the \citet{Smartt2009} mass estimates, the different effective
survey volumes for SN~II-Ps
and OTs, and the lack of a secure progenitor mass for SN~2008S, we
cannot yet draw a firm conclusion.

\subsection{Limitations of the Method}
\label{sec:limitations}

The work above demonstrates that one can obtain accurate precursor
masses even when the actual precursor is undetected.  Although the
images we used were taken several years before the transient, there
was nothing in the analysis that required direct identification of
the precursor.  Instead, the mass was derived through ``guilt by
association'' using the properties of the underlying burst of star formation.  In
principle, one could use this method to derive masses of precursors of
every transient or SN in the historical record.
In practice, however, this method is somewhat more limited.  

\subsubsection{Crowding}

Recovery
of recent SFHs requires CMDs of individual main-sequence and
helium-burning stars.  However, images of individual stars are blurred
together, with the crowding being more severe at larger distances, in
high surface brightness regions, and/or for fainter
stars (which are more numerous).
Even with \emph{HST}'s resolution, crowding quickly limits the
ability to measure fluxes for all but the brightest stars beyond
$\sim 4$--5 Mpc.  The crowding limit can be reached many magnitudes
brighter than one would expect for a given telescope aperture and
exposure time.  This limitation is less of a problem for stars on the blue and red
helium burning sequences, which are far more luminous than main-sequence
stars of comparable masses.  However, these sequences are only well-populated 
after $\sim 25$ Myr, making them useful only for low mass precursors
($M \lesssim 9$ \Msun).  Thus, crowding is a significant limitation on
the applicability of this method, at least until the launch of
telescopes that can surpass \emph{HST}'s angular resolution in the
optical.  Advances in infrared space telescopes may not bring
much progress, given the faintness of main-sequence stars in the
infrared.

\subsubsection{Depth}

Even when crowding is not significant for a main
sequence star of a given stellar mass, the SFH method requires depths
that are a magnitude or
more below the main-sequence turnoff of the stellar masses of
interest.  The recent star formation is largely constrained by the
luminosity function of the main sequence, so that the behavior on the  
more
sparsely populated upper main sequence is forced to be consistent with
the numbers of stars on the lower main sequence.  Thus, bursts are
best constrained when the main sequence is well-measured
substantially below the turnoff.

\subsubsection{Very Massive Stars}

While the modeling in this paper yielded excellent limits on the mass
of this particular precursor, we anticipate larger uncertainties for
higher mass precursors, for several reasons.  First, the optical
luminosity of a very massive main-sequence star is a tiny fraction of
its bolometric luminosity, making optical CMDs poor constraints on the
masses of the most massive stars.  The masses of the most massive main
sequence stars can only be accurately determined through spectroscopy
\citep{massey03}, given that stars with $T>30,000$K have essentially
identical optical and near-UV colors.  Thus, while using SFHs to
measure precursor masses may work for up to 20--30 \Msun, it is less
likely to be useful for higher masses.  

Second, the very youngest,
most massive stars are likely to remain shrouded within their natal
molecular clouds to some extent.  If these stars have not emerged from
their dust cocoons by $\sim 5$ Myr (the turnoff age of a $\sim 25$
\Msun\
star), then their measured luminosities and colors are likely to be
significantly affected by dust, causing the main sequence to broaden
to the red and to shift to lower luminosities.  SFH recovery codes
typically take dust into account statistically by including a
``differential reddening'' term to help match the width and position
of the main sequence.  However, this correction cannot compensate for
stars that are simply absent due to very high dust extinction;
observations in the Magellanic clouds suggest that this effect is
significant for ages less than 2 Gyr \citep{massey95}.  Dust can also
be corrected for on a star-by-star basis, provided that imaging on the
UV side of the Balmer break is included \citep{romaniello02}.  

Third,
very massive stars are likely to be found within dense stellar
clusters that have not yet experienced significant cluster
dissolution.  The timescale for this dissolution is $\sim 10$ Myr, such
that 50\% of stars with mass 15 \Msun\ are likely to still be bound
into clusters \citep{Lada2003,Bastian2005}.  These clusters have
higher than average crowding, and thus stars within them are prone to
larger photometric errors and brighter limiting magnitudes.  

Finally,
the upper main sequence is always less populated than the lower,
making it statistically possible that the youngest apparent
``turnoff'' in the main sequence appears to be older than the true
turnoff, due to the sparse sampling of the high mass end of the IMF.

\subsubsection{Region Size}
\label{sec:region}

An additional concern with the method is the choice of region size.
If the radius within which stars are selected is too small, it is
possible that stars will have dispersed from an unbound cluster to
beyond this radius.  However, the dispersal timescale is tens of Myr,
as the stars are expected to disperse with the typical cluster
velocity dispersion of a few km~s$^{-1}$ \citep{Bastian2006}.  In the
$5\arcsec$ (50 pc) region we selected, if the cluster velocity
dispersion is 1 km~s$^{-1} = 1$ pc Myr$^{-1}$ (a reasonable value for
a small cluster), stars would still be contained within the region
after 50 Myr, which is longer than the estimated timescale of star
formation for this cluster.  For a larger cluster with a velocity
dispersion of 3 pc Myr$^{-1}$, the stars would remain within 50
pc for 17 Myr, still longer ago than the beginning of the star
formation event we see in this region.

Choosing a larger radius would ensure that all stars from the cluster
are included; however, a larger radius also increases contamination
from surrounding unrelated stars.
We find that using a radius of $7.5\arcsec$
instead of 5 (2.25 times the area enclosed) results in a less
well-defined burst, probably because we are enclosing stars slightly
older than the population which gave rise to the transient.

\subsubsection{Choice of Stellar Evolution Models}
\label{sec:models}

We have used the isochrones from the Padova group
\citep{Marigo2008} in our analysis, and the errors reported assume
that those models are correct.  However, additional systematic errors
may be introduced by uncertainties in the models.  
\citet{Gallart2005} provide an in-depth discussion of the
discrepancies between different models and their effect on deriving
SFHs using CMD fitting, and they conclude that the uncertainties seem
to be smallest for young stars, which is reassuring for studies of
transient progenitors.
To assess the errors in more detail, we
compare the Padova models with the currently available Geneva models
\citep{Lejeune2001}.  Both sets of models include the effects of
convective core overshooting, but not rotation.  The Padova models
include thermal pulses on the AGB, while the Geneva models do not.

\begin{figure}
\plotone{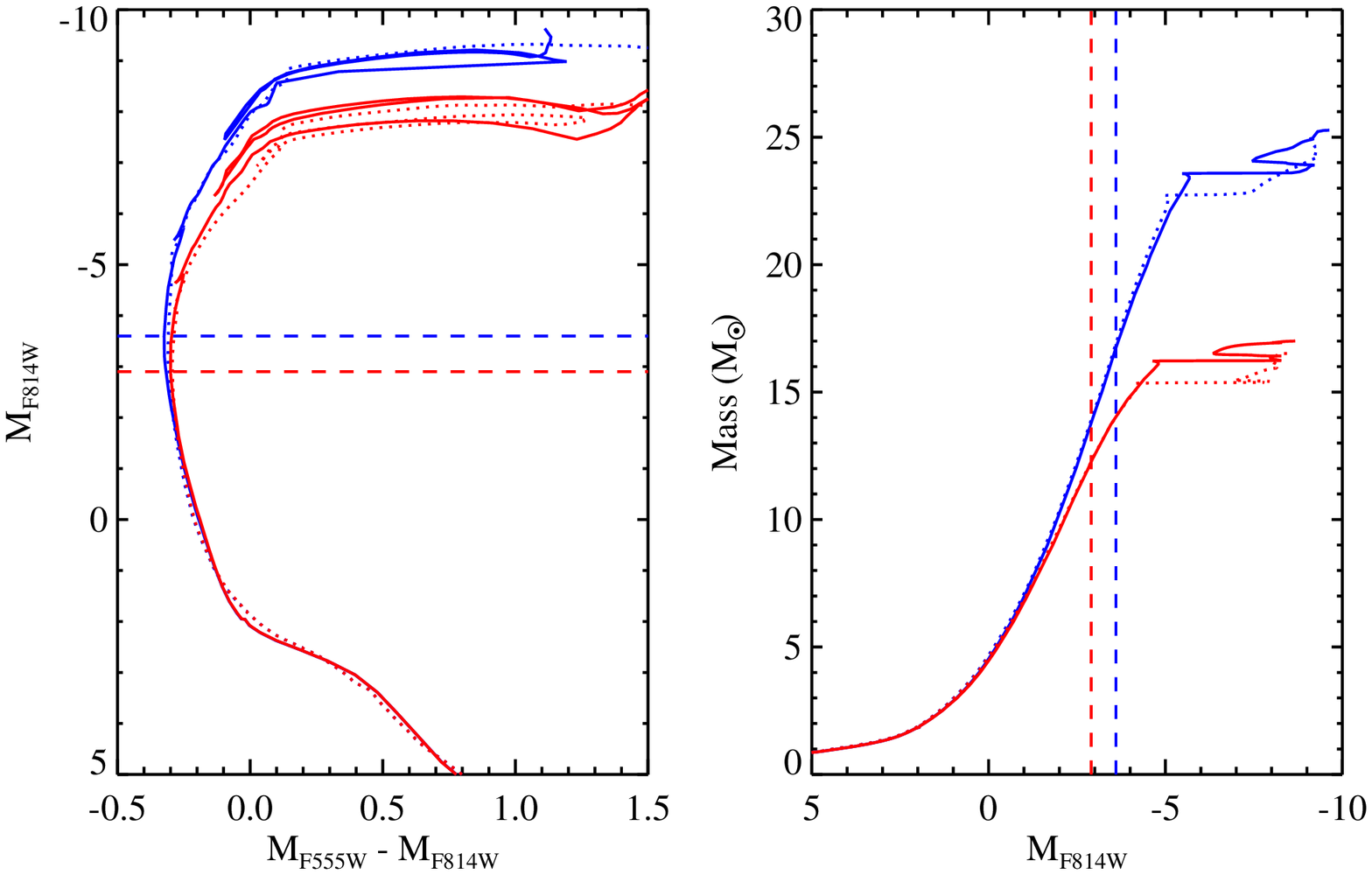}
\caption{\label{fig:iso_compare}
  Comparison of the Padova \citep[][solid lines]{Marigo2008} and Geneva
  \citep[][dotted lines]{Lejeune2001} isochrones for
  $Z=0.008$ and ages of 8 Myr (blue) and 13 Myr (red).  The left panel
  shows the isochrones on a CMD, while the right panel shows stellar mass
  as a function of absolute magnitude.  The magnitude of the
  main-sequence turnoff is
  highlighted with a dashed line for each age.
  The maximum masses shown by the endpoints of the isochrones cover a
  range of 16--25 \Msun.
  The two different isochrone sets agree well for main sequence stars,
  but show significant differences in their post-main-sequence
  evolution.}
\end{figure}

Figure~\ref{fig:iso_compare} shows isochrones from the two
models for ages 8 Myr and 13 Myr, the flanking ages of the star
formation event we found for \ot.  Metallicity is $Z=0.008$ ($\mh =
-0.4$).  The left panel plots the isochrones for both models
for each age, with the main-sequence turnoff magnitude marked.  The
models are very similar up to the turnoff, and diverge for more
massive stars.  We are especially concerned with the mapping between
age and stellar mass, since we use this relationship to go from an age
estimate to a stellar mass estimate for the \ot\ progenitor, so we
plot mass as a function of absolute magnitude for the same isochrones in the right panel of
Figure~\ref{fig:iso_compare}.  For each age, the mass-luminosity
relationship is identical between the models up until the main-sequence turnoff, but there are discrepancies above the turnoff.
Since the age of a young stellar population is primarily determined by
the main sequence, we expect that differences in the models related to
the treatment of post-main-sequence phases will not significantly
affect our results.

A potentially significant effect not addressed in the previous
comparison is rotation.  
For solar metallicity and an initial rotation
velocity of 200 km
s$^{-1}$, the age estimate for a star of a given luminosity would
increase by about 25\% from the non-rotating model \citep{Meynet2000}.
Hence, including rotation in the models would increase our age
estimate for the star formation event to 10--16 Myr and the upper mass
limit
for the turnoff
would decrease to 15 \Msun.  The lower mass limit 
for the turnoff
changes from 12.3 \Msun\ to 11.9 \Msun.  Hence, even including the
effects of rotation at this velocity, our
turnoff
mass range of 12--17 \Msun\ is still
valid.  Rotation
speeds in excess of 200 km s$^{-1}$ could lead to smaller possible
masses for the progenitor.
\citet{Vazquez2007} show that the decrease in main sequence lifetime
also holds for stars at $Z = 0.008$ (the approximate metallicity of
the region around the transient), but isochrones for this
metallicity are only available for masses $>30$ \Msun.

\subsubsection{Ambiguous Bursts}

The final possible limitation is that Nature may not always be as kind
as she has been in this particular instance.  While \ot\
came from an unambiguous isolated burst, some transient
phenomena may be associated with regions containing multiple bursts at
a range of ages, making the association of mass ambiguous.  This
effect is more likely to be pronounced in more distant galaxies,  
where a typical
aperture size of a few arcseconds encompasses a larger physical area
of the disk, and thus is more likely to contain more than one
dissolving open cluster.  On the other hand, the analysis of star
formation regions in M81 ($D \sim 3.7$ Mpc) presented in
\citet{Gogarten2009} uses $\sim 27 \arcsec \approx 500$ pc regions, and
finds bursts that are well localized in time up to 100 Myr.

\subsection{Benefits of the Method}

In spite of the above limitations, there are a number of clear
benefits to the procedure we have demonstrated in this paper.  The
first and most obvious one is the removal of the need for precursor
imaging in determining the mass.  Instead, targeted observations after
the event are sufficient, and can be optimized for recovery of the
SFH, through proper choices of filters and observing strategy.
Second, unlike in precursor imaging, one does not need
accurate astrometry.  Since the SFH recovery uses stars within a few
arcsecond wide aperture, it is not required that the SN or transient
be localized
to better than half the width of the selection aperture.  Third, given
the uncertainties in the late stages of stellar evolution, it is
perhaps easier to derive an accurate mass from the bulk properties of
the better-understood main- and helium-burning sequences, rather than
from the luminosity and SED of a star on the
brink of explosion.
Finally, the relationship between progenitor mass and SN type as
predicted by stellar evolution models relies on assumptions about mass
loss during late stage stellar evolution.  Contributions to the
existing catalog of precursor masses using the SFH technique will
improve the statistics in comparing observations with stellar
evolution theory, and may even give more reliable initial mass
estimates than can be determined from an individual star.

\section{Conclusions}
\label{sec:conclusions}

In this paper we have demonstrated a technique for measuring the
masses of precursors of luminous transient events, even when no
pre-event imaging exists.  By using standard tools to recover the SFH
from stellar populations, we find that \ot\ originated from stars
formed in a burst between 8 and 13 Myr ago with a certainty of 70\%.
The current turnoff mass associated with this burst is 12--17 \Msun\.
Assuming the transient is due to an evolving massive star, then the
mass of the precursor must be higher than this turnoff mass, but less
than the 16--25 \Msun\ mass limit above which stars of this age have
exploded.  The resulting mass range of 12--25 \Msun\ agrees quite well with estimates
of 10--15 \Msun\ by \citet{Bond2009} and 10--20 \Msun\ by
\citet{Berger2009}.  This technique therefore shows great promise for
significantly expanding the number of SNe and transients with reliable
precursor masses.

The SFH method presented here for determining precursor masses is
largely complementary to direct precursor imaging, as each method has
benefits and limitations.
At present, the limits of the SFH technique are largely observational.
The angular resolution of \emph{HST} currently restricts this  
technique to
well within $D<10$ Mpc.  
In a future paper (Gilbert et al.\ 2009, in preparation), we will
present an application of the method to all SNe and transients for
which sufficiently deep data exists in the \emph{HST} archive.

With future space-based optical
telescopes one can certainly push the technique to much larger
distances, greatly expanding the volume of candidate SNe.  The more
intensive monitoring of nearby galaxies can also significantly
increase the number of SNe for which this technique can be used;
thankfully such monitoring is currently underway \citep[e.g.,][] 
{Leaman2008,DiCarlo2008}.  
The main benefits of the method are that imaging can be done after a
transient event, even without accurate astrometry, and that fitting
the entire main sequence of a star formation event may provide more
reliable mass estimates.

In closing, we note that precursor imaging is still highly desirable.
The technique we have employed gives only a constraint on the
main-sequence mass of the star that eventually erupted.  However, it
gives
no constraint on the exact phase of stellar evolution that the star
was in immediately before the eruption.
Precursor imaging can yield much more information about a star than
its mass alone (e.g., color, magnitude).
Given the impact of the
unexpected phase of the SN~1987A precursor (i.e., that it was a blue,
rather than red, supergiant at the time of explosion) on our
understanding of core-collapse SNe, any additional information on the
precursor is likely to be highly significant.

\acknowledgements We thank Adam Burrows, Luc Dessart, Christian Ott,
and Nathan Smith for helpful discussions.  Leo Girardi provided the
stellar evolution models used in this paper as well as a discussion of
their uncertainties.  We also thank the anonymous referee for comments
which significantly improved the paper.  Support for this work was
provided by NASA through grant GO-10915 from the Space Telescopes
Science Institute, which is operated by the Association of
Universities for Research in Astronomy, Inc., under NASA contract
NAS5-26555.  S.M.G. was partially supported by NSF grant CAREER AST
02-38683.  J.J.D.\ was partially supported as a Wyckoff Faculty
Fellow.  In addition, J.W.M.\ is supported by an NSF Astronomy and
Astrophysics Postdoctoral Fellowship under award AST-0802315.

{\it Facilities:} \facility{HST (ACS)}


\end{document}